\documentclass{svproc}
\pdfoutput=1

\usepackage{epsfig}
\usepackage{hyperref}

\newcommand{\tit}{\subsection}
\newcommand{\rf}[1]{(\ref{#1})}
\renewcommand{\d}{{\rm d}}

\newcommand{\om}{{\omega}}

\newcommand{\hg}{\hat g}
\newcommand{\tr}{{\rm tr}\,}

\renewcommand{\!}{\negthinspace}
\newcommand{\LA}{\left\langle}
\newcommand{\RA}{\right\rangle}
\def\e{{\,\rm e}}
\def\be{\begin{equation}}
\def\ee{\end{equation}}
\def\beq{\begin{equation}}
\def\eeq{\end{equation}}
\def\bea{\begin{eqnarray}}
\def\eea{\end{eqnarray}}
\newcommand{\eq}[1]{Eq.~(\ref{#1})}
\newcommand{\eps}{\varepsilon}
\newcommand{\non}{\nonumber \\*}

\newcommand{\vp}{\varphi}
\newcommand{\p}{\partial}
\newcommand{\bp}{\bar\partial}

\newcommand{\tbl}[1]{#1}
\newcommand{\tre}[1]{#1}
\newcommand{\tbr}[1]{#1}

\newcommand{\tma}[1]{#1}
\newcommand{\tcy}[1]{#1}

\usepackage{url}

\begin{document}
\mainmatter              
\title{Strings from Nambu-Goto to Polyakov and back}
\titlerunning{Strings from Nambu-Goto to Polyakov and back}  
%
\author{Yuri Makeenko}
\authorrunning{Yuri Makeenko} 
%
\tocauthor{}
\institute{NRC ``Kurchatov Institute''\/-- ITEP, Moscow \\[2mm]
\email{makeenko@itep.ru}}

\maketitle              

\begin{abstract}
I discuss the recent progress in  bypassing the KPZ barrier
for the existence  of nonperturbative bosonic strings in $1<d<25$.
I~consider string anomalies which emerge from higher terms of the DeWitt-Seeley expansion as
 $\eps \times \eps^{-1}$ with $\eps$ being a UV cutoff. I show they give a nonvanishing contribution to
the central charge and  to 
the string susceptibility, telling the Nambu-Goto and Polyakov strings
apart. 
I describe an exact solution of the emerging four-derivative two-dimensional conformal theory
and its relation to minimal models.\\[2mm]
Talk at ``Gravity, Strings and Fields: A Conference  in Honour of Gordon Semenoff'', Montreal July 24--28, 2023.
\end{abstract}


\label{sec:intro}
\section{Brief introduction}

This Talk is about  the problem of the existence  of nonperturbative bosonic strings in $d=4$
space-time dimensions inherited from 1980's.
The quantity of interest is the string susceptibility index $\gamma_{\rm str}$ defined through
the number of surfaces of large area $A$ by
\be
\LA \delta \left(\int \sqrt{g} -A \right) \RA \propto A^{\gamma_{\rm str}-3} \e^{C A}.
\ee
For  closed {Polyakov's string} it was calculated by
Knizhnik-Polyakov-Zamolodchi\-kov~\cite{KPZ88} (KPZ) and David~\cite{Dav88}, 
Distler-Kawai~\cite{DK89} (DDK) using the technique of conformal field theory (CFT).
The result
\be
\gamma_{\rm str}=2+(\tcy{h}-1)\frac{12 +(d_+ +d_-)/2-d+\sqrt{(d_+ -d)(d_- -d)}}{12}
\quad ~(\hbox{genus}~h)
\label{mygstr}
\ee
with $d_+=25$, $d_-=1$ was real only if $d\leq 1$ or $d\geq 25$ but not in $d=4$
(the KPZ barrier).

I argue in this Talk  that a possible solution of the problem  may be
that the barrier is shifted to $d_->4$ for the Nambu-Goto string.
Such a situation is realized for the four-derivative CFT I shall momentarily describe.

\section{From the Nambu-Goto string to four-derivative CFT}

\subsection{Nambu-Goto versus Polyakov strings}

The action of the {Polyakov string} is
{quadratic} in the target-space coordinate
$X^\mu$ and has an {independent} metric tensor $g_{ab}$:
\be
{\cal S}=
\frac{K_0}2 \!\int \d^2\omega\, \sqrt{g}g^{ab}  \partial_a X \cdot \partial_bX,\qquad
K_0=\frac 1{2\pi\alpha'}.
\ee
{The action of the Nambu-Goto string} can also be written likewise by introducing 
 {the (imaginary) Lagrange multiplier} $\lambda^{ab}$:
\be
K_0 \!\int \d^2\omega\,\sqrt{\det \partial_a X \cdot \partial_bX}=
K_0 \!\int \d^2\omega\,\sqrt{ g}  
+\frac{K_0}2 \!\int \d^2\omega\, \lambda^{ab} \left( \partial_a X \cdot \partial_bX -g_{ab}
\right).
\ee
 The two string formulations are equivalent classically when 
 $g^{\rm cl}_{ab}=\p_a X \cdot \p_b X$ coincides
 with the induced metrics  (and  $\lambda^{ab}_{\rm cl}=
 \sqrt{g}g^{ab}$).
 I shall return soon to the general argument from the book~\cite{Pol87} about the equivalence.

Let us  split
$X^\mu=X^\mu_{\rm cl} +X^\mu_{\rm q}$
and do the {Gaussian} path
integral over the quantum part $X^\mu_{\rm q}$:
\be
S_{\rm ind}= K_0 \int \d^2\omega\,\sqrt{ g} 
+\frac{K_0}2 \int \d^2\omega\, \lambda^{ab} \left( \partial_a X_{\rm cl } \cdot \partial_bX_{\rm cl } -g_{ab} \right) + \frac{d}{2}  \tr \log {\cal O},
\label{ind}
\ee
where the
operator ${\cal O}= -\frac1 {\sqrt{g} } \partial _a \lambda^{ab} \partial_b$ 
reproduces  the Laplacian $\Delta$ for
$\lambda^{ab}_{\rm cl}$.
An {additional}  {ghost determinant} emerges as usual
in the conformal gauge.
The action~\rf{ind} is called 
{induced} (or {emergent}). For {smooth} fields it coincides with the {effective} 
action.



Two-dimensional  determinants diverge and are to be {regularized}.
For Schwin\-ger's {proper-time} regularization
the integrals over $\tau$ are simply cut from below at $a^2=1/4\pi \Lambda^2$.
Alternatively, one can use 
{Pauli-Villars'} regularization  
when
\be
 \tr\log{\cal O}\big|_{\rm reg}= -
\int_{0}^\infty \frac{\d \tau}{\tau} \,\tr \e^{-\tau{\cal O}}
\left(1-\e^{-\tau M^2}\right)^2, \quad ~~\Lambda^2 = \frac{M^2}{2\pi}\log 2
\ee
is {convergent} at finite {regulator mass} $M$ and divergent as $M\to\infty$.

For  {Pauli-Villars'} regularization Feynman's diagrammatic  technique applies and 
the det's can be {exactly}
computed for certain metrics by the
 {Gel'fand-Yaglom} technique  to compare with the {DeWitt-Seeley} expansion
of $\LA \om |\e^{-\tau {\cal O}} |\om\RA$ which starts with the term  $1/\tau$
in  two dimensions.
For $\tau\sim \Lambda^{-2}$ the higher terms are suppressed as 
powers of $R/\Lambda^2$
for smooth fields but revive if they are not smooth.
Higher-derivative terms 
are related by simple formulas~\cite{Mak23a} for the two regularizations. 

\subsection{Emergent action ${\cal S}[\vp,\lambda]$}

Integrating out $X^\mu_q $ we get (a part of) the {emergent action}
\be
\frac {d}2\,{\rm tr}\ln \left[-\frac1{{\rho}}\partial _a \lambda^{ab}\partial_b
\right] _{\rm reg}~=~
\sum_n \frac 1n \!\!\!\!
\unitlength=.5mm
\linethickness{1.6pt}
\begin{picture}(37.00,14.50)(1,18.6)
\put(20.00,20.00){\circle{10.00}}
\multiput(16.46,23.535)(-2.00,3.33){3}{
  \bezier{28}(0.00,0.00)(-1.333,0.333)(-1.00,1.665)}
\multiput(15.46,25.20)(-2.00,3.33){2}{
  \bezier{28}(0.00,0.00)(0.333,1.167)(-1.00,1.665)}
\multiput(23.535,23.535)(2.00,3.33){3}{
  \bezier{28}(0.00,0.00)(1.333,0.333)(1.00,1.665)}
\multiput(24.535,25.20)(2.00,3.33){2}{
  \bezier{28}(0.00,0.00)(0.333,1.167)(1.00,1.665)}
\put(20.00,30.00){\makebox(0,0)[cc]{$\cdot $}}
\put(23.00,29.50){\makebox(0,0)[cc]{$\cdot $}}
\put(17.00,29.50){\makebox(0,0)[cc]{$\cdot $}}
\put(11.00,15.50){\makebox(0,0)[cc]{$\cdot $}}
\put(9.80,18.50){\makebox(0,0)[cc]{$\cdot $}}
\put(12.90,13.00){\makebox(0,0)[cc]{$\cdot $}}
\put(27.10,13.00){\makebox(0,0)[cc]{$\cdot $}}
\put(29.00,15.50){\makebox(0,0)[cc]{$\cdot $}}
\put(30.20,18.50){\makebox(0,0)[cc]{$\cdot $}}
\multiput(20.00,15.00)(0.00,-4.00){2}{
  \bezier{28}(0.00,0.00)(-1.00,-1.00)(0.00,-2.00)}
\multiput(20.00,13.00)(0.00,-4.00){2}{
  \bezier{28}(0.00,0.00)(1.00,-1.00)(0.00,-2.00)}
\end{picture}
\ee
where the wavy lines correspond to fluctuations $\delta \lambda^{ab}$
or $\delta g_{ab}$ about the ground state.  
In the conformal gauge 
\be
g_{ab}=\hat g_{ab}\e^\vp 
\label{confog}
\ee
we find  the following contribution from  $X^\mu$  and its Pauli-Villars' regulators to the emergent 
action \cite{Mak23c}:
\bea
{\cal S}_X[\vp, \lambda] 
&= &\frac d2 \int \left[ -\frac {\sqrt{\hg}\e^{\vp}\Lambda^2}{ \sqrt{\det{\lambda^{ab}}}}
+\frac1{48\pi}\left( \sqrt{\hg}\vp \hat\Delta\vp +
\lambda^{ab} \hg_{ab}\hat \Delta \vp+2
\lambda^{ab}\nabla_a\p_b  \vp  \right)\right. \non &&\hspace*{.9cm}\left.
+\frac{\sqrt{\hg}\e^{-\vp}}{160\pi M^2}  (\hat\Delta \vp)^2 \right] 
+ {\cal O}(M^{-4}),\qquad \quad
\Lambda^2=\frac{M^2}{2\pi}\log 2
\label{SXb}
\eea
with $\nabla_a$ being the covariant derivative for the metric tensor 
$g_{ab}$ given by \eq{confog} and 
$\hat \Delta=\e^{\vp}\Delta$ is the Laplacian for the metrics tensor $\hg_{ab}$.
The ghost contribution to the emergent action is as usual.

Expanding to the quadratic order in $\delta \lambda^{ab}$, we see that its mass 
squared $\sim\!\! \Lambda^2$, so $\delta\lambda^{ab}$  is
expected to decouple~\cite{Pol87}. 
Rather surprisingly, its private life which occurs at the distances $\sim \Lambda^{-1}$ is 
nevertheless observable~\cite{Mak21}
like anomalies in quantum field theory (QFT). I shall now start describing this issue.

\subsection{Four-derivative Liouville action ${\cal S}[\vp]$}

Path-integrating out $\delta\lambda^{ab}$ in \rf{SXb} and integrating 
the emergent action by parts, we obtain 
modulo boundary terms the four-derivative action~\cite{Mak21}
\be
{\cal S}[\vp]=\frac 1{16 \pi b_0^2} \int \!\sqrt{g} \left[g^{ab} \p_a \vp \p_b \vp  +2\mu_0^2
+\eps  R \left(R+\tbl G g^{ab}\,\partial_a \vp\partial_b \vp \right)\right]\!,~~ b^2_0=\frac 6{26-d}
\label{inva}
\ee
where $\vp=-\Delta^{-1} R$ becomes a local field in the conformal gauge~\rf{confog}
and $\eps\propto M^{-1}$ is the worldsheet cutoff.  
 Only the two four-derivative terms shown in \eq{inva} are independent.
The first additional term $R^2$  appears already
for Polyakov's string from the DeWitt-Seeley expansion of the heat kernel but the 
second  one with $G\neq 0$ is specific to the Nambu-Goto string.
We have dropped terms of higher orders in $\eps$ (to be discussed below). 


Of course the higher-derivative terms vanish classically for smooth metrics with 
$\eps R\ll1$, reproducing the usual Liouville action. However, the
quartic derivative provides both  an ultraviolet 
(UV) cutoff and also an \tre{interaction} whose
{coupling} constant is $\eps$.  We thus encounter 
 uncertainties like powers of $\eps \times \eps^{-1}$ so the higher-derivative terms  revive \tma{quantumly}. 
 In other words the
\tbr{smallness of $\eps $ is compensated by a change of  the metric (the shift of $\vp$)}
what is specific to the theory with diffeomorphism invariance.

 The described procedure looks like an appearance of anomalies in QFT. We may expect for this reason that possible yet higher-derivative terms 
 will not change the results. Such a universality
 at $G=0$ was argued in \cite{Mak21} and proven in~\cite{Mak23c}.

\section{Conformal invariance of four-derivative Liouville action}

\tit{``Improved'' energy-momentum tensor}  

To apply the methods of CFT to the four-derivative 
Liouville theory~\rf{inva}, we first derive its energy-momentum tensor $T_{ab}$.
Naively, one simply substitutes
the decomposition~\rf{confog} into the action~\rf{inva} and takes the variation with
respect to $\hg_{ab}$. But this would give only a part of $T_{ab}$ associated with the
minimal coupling of $\vp$ to the background gravity $\hg_{ab}$. The correct procedure
is to use 
\be
\sqrt g R= \sqrt{\hat g} \left(q \hat R - \hat \Delta \vp \right) \qquad
(q=1 ~~~\hbox{classically})
\label{Rshift}
\ee
under the decomposition~\rf{confog}.
This results in an appearance of the additional terms
 in the action, resulting in a non-minimal interaction of $\vp$ with background gravity.
 These additional terms are crucial for the Weyl covariance of the four-derivative action~\rf{inva}.
 
 The  energy-momentum tensor associated with the minimal interaction is conserved thanks to the 
 classical equation of motion but not traceless owing to the presence of the dimensionful
 parameters like $\mu_0^2$ and $\eps$. 
 The ``improved'' energy-momentum tensor~\cite{DJ95} that
involves additional terms, coming from the non-minimal interaction with background gravity,
is conserved and {\em traceless}\/ thanks to the 
classical equation of motion for $\vp$ in spite of the presence of 
dimensionful $\mu_0^2$ and $\eps$.
 This is a general property because in the conformal gauge \rf{confog}
we have
\be
T^a_a\equiv\hat g^{ab} \frac{\delta {\cal S}[g]}{\delta \hat g^{ab} }= -
\frac{\delta {\cal S}[g]}{\delta \vp },
\label{tra}
\ee
where the left-hand side represents the trace of the 
``improved'' energy-momen\-tum tensor
while the right-hand side represents the classical equation of motion for $\vp$.
The four-derivative action \rf{inva}
 is thus conformal invariant for flat backgrounds.
 
 Before explicitly writing the energy-momentum tensor let me mention the renormalization of the parameters $b_0^2$ in \eq{inva} and $q$ in \eq{Rshift} in the quantum case:
 $b_0^2\to b^2=b_0^2+{\cal O}(b_0^4)$ and $q_0=1\to q=1+{\cal O}(b_0^2)$.
 This is like it was pointed out by DDK~\cite{Dav88,DK89} for the usual Liouville action.
 We can think of the renormalization as a result of path-integrating out the regulator fields.
 Accounting for the renormalization, we write for the $T_{zz}$ component~\cite{Mak22}
\bea
\lefteqn{ T_{zz}=}\non &&-\frac1{4b^2} \big[ (\p \vp)^2 -2\eps \p \vp \p \Delta \vp 
-G \eps (\p \vp)^2 \Delta \vp+ 4G\eps\p\vp \p(\e^{-\vp}\p \vp \bp \vp)  
 +  G \eps\partial(\p \vp \Delta\vp)\big] \non&&+\frac q{4b^2}\left[ 2\p^2 (\vp - \eps\Delta \vp) 
+ 4G \eps \p^2(\e^{-\vp}\p \vp \bp \vp)-
  G \eps \frac 1{\bp}\p ^2 (\bp \vp \Delta\vp)\right].
  \label{qTzz} 
\eea
Notice the appearance of the nonlocal (last) term as inherited from the nonlocality of the covariant
action~\rf{inva}.

\tit{DDK for the four-derivative Liouville action} 

Given $T_{zz}$ it is possible to  compute \'a la DDK the conformal weight and the central charge at one loop 
about the classical ground state.
The operator products $T_{zz}(z) \e^{\vp(0)}$ and  $T_{zz}(z) T_{zz}(0)$ 
are given at one loop by the diagrams depicted in Fig.~1.
\begin{figure}[t]
\begin{center}
\vspace*{.2cm}
\includegraphics[width=10cm]{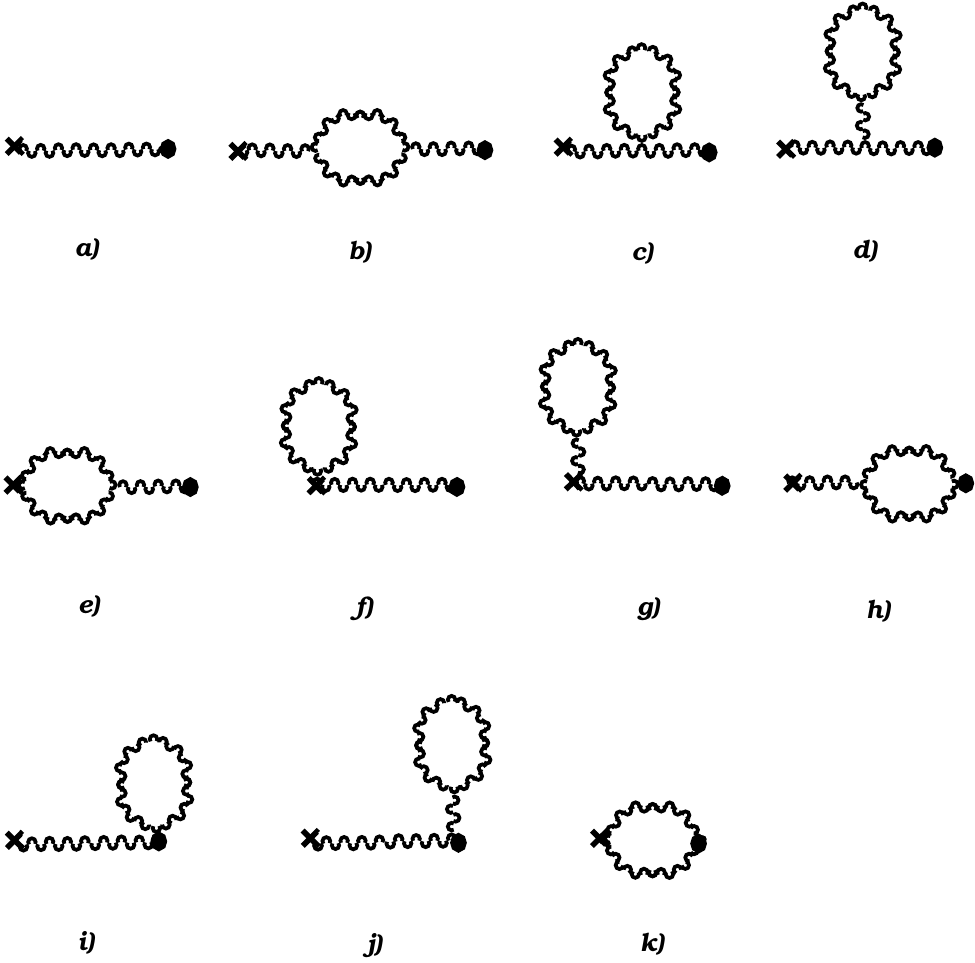} 
\caption{One-loop diagrams contributing to  the operator-product expansion of
$T_{zz}(z) \e^{\vp(0)}$ or $T_{zz}(z) T_{zz}(0)$. The points $z$ and 0 are depicted 
by the cross and the dot, respectively.}
\end{center}
\label{fi:2}
\vspace{-5mm}
\end{figure}
The diagrams a) to d) renormalize $b_0^2\to b^2$ 
in the propagator and e) to g)
renormalize $q_0=1\to q$.
The diagram k) contributes $-b^2$ 
to the conformal weight of $\e^{\vp(0)}$ which equals 1, so we have as usual
\be
 1=q-b^2.
 \label{DDK1} 
\ee

For the central charge the
diagrams a) to d) contribute $6 q^2/b^2$  as usual.
The diagram k) contributes the usual $1$  but now
the nonlocal term in \eq{qTzz}  revives and gives additional $6 G q$.
There is also a logarithmically divergent term 
which does not appear if $T_{zz}$ is normal-ordered.
Adding the three terms we find
\be
c^{(\vp)}=\frac{6q^2}{b^2} +1 +6 G q 
\label{DDK2}
\ee
at one loop. As will be shown below, actually Eqs.~\rf{DDK1} and \rf{DDK2} are {\em exact}.

\section{{\bf Algebraic check of DDK}}

\tit{Covariant Pauli-Villars' regularization}  

To describe the one-loop renormalization in QFT, we add
{Pauli-Villars' regulators}: \tbr{Grassmann} $Y$, $\bar Y$ (of mass squared $M^2$) and 
\tbr{normal} $Z$ (of mass squared $2M^2$). This regularizes all involved divergences
including those in tadpoles.
To simplify formulas I keep below only one $Y$ which is enough to compute finite parts.
The regulator action  reads
\be
{\cal S}_{\rm reg} =\frac 1{16\pi b_0^2}\int \sqrt{g}\left[ g^{ab}\partial_a  Y \partial_b Y +M^2 Y^2+
\eps (\Delta Y)^2 +G\eps g^{ab}\partial_a  Y \partial_b Y R  \right]. 
\ee

The regulators make a contribution to the energy-momentum which is quad\-ra\-tic in the regulator fields
and local.  The total energy-momentum tensor is
conserved and traceless (!) as before in spite of the mass.
 We thus expect conformal invariance  to be maintained quantumly what can be
 explicitly checked by the one-loop and partially two-loop calculations.
 
\tit{Explicit one-loop results}  

The renormalization of  $b^2$  comes from the usual one-loop diagrams including tadpoles:
\be
\frac{1}{b^2}=\frac1{b_0^2}-
\left(\frac 16 -4 +A +2 G \int \d k^2 \frac \eps{(1+\eps k^2)}-
\frac 12 G A\right)+{\cal O}(b_0^2) .
\ee
Here $A(\eps M^2)\sim \eps M^2$ ($\eps M^2\gg1$) is the contribution of the tadpole. 

The analogous one-loop renormalization of $q/b^2$ in $T_{zz}$ reads
\be
\frac q{b^2}=\frac{1}{b_0^2}-\frac 16 +2 -\frac 12 A -\frac 12 G 
- G \int \d k^2 \frac {\eps }{(1+\eps k^2)} +\frac 14 G A 
\ee
or, multiplying by $6 b^2$,
\be
\frac {6q^2}{b^2} =\left(\frac q{b^2}\right)^2\times 6 b^2
=\frac{6}{b_0^2} - 1 -6 G +{\cal O}(b_0^2).
\ee
This precisely confirms~\cite{Mak21} the addition to the central charge in \eq{DDK2}
obtained by the CFT technique.

\section{The method of singular products\label{s:dim}}

For a general higher-derivative action  $S[\vp]$  
one sees tremendous cancellations when computing the operator products
 $T_{zz}(z)\e^{\vp(0)}$ or  $T_{zz}(z)T_{zz}(0)$. To account for the cancellation,
 it is convenient to write the generator of 
the conformal transformation as
\be
\hat \delta_\xi 
=\int_{D_1} \d^2 z\left( q\xi'(z) \frac {\delta}{\delta \vp(z)}
 +\xi (z)\p \vp(z)\frac {\delta }{\delta \vp(z)}\right),
 \label{hatdel}
 \ee
where the domain $D_1$ includes the singularities of $\xi(z)$ leaving outside possible singularities of the string of
functions $X(\om_i)$ on which $\hat \delta_\xi $ acts. 
Actually, the form of $\hat \delta_\xi$ on the right-hand side of \eq{hatdel} is primary.
Its advantage over the standard
one based on $T_{zz}$ is that it accounts for the cancellations  in the quantum case, 
while there are subtleties associated with singular products. 

\tit{Conformal weight}

It is easy to reproduce
\be
\hat \delta_\xi \e^{\alpha\vp(\om)} \stackrel{{\rm w.s.}}=
(q \alpha-b^2 \alpha^2)\xi'(\om)\e^{\alpha\vp(\om)} +\xi(\om)\p\e^{\alpha\vp(\om)}
\label{29}
\ee
 for the quadratic action by \rf{hatdel} via the singular products:
\bea
\hat \delta_\xi \e^{\alpha\vp(\om)  }
 &\stackrel{{\rm w.s.}}=&
q\alpha \xi'(\om)  \e^{\alpha\vp(\om)} +\int_{D_1}\!\d^2 z \, \alpha^2\xi(z) 
\big\langle  \p\vp(z) \vp(\om)\big\rangle \delta^{(2)}(z-\om)  \e^{\alpha\vp(\om)}\non &&
 +\alpha \xi(\om)   \p\vp(\om)\e^{\alpha\vp(\om)}. 
\label{32}
 \eea
 where the equality is understood in the week sense, i.e.\ under path integrals.
 The most interesting is the second term on the right-hand side, where the singular product 
 equals~\cite{Mak22c}
\be 
\int_{D_1}\!\d^2 z \, \xi(z) 
\big\langle  \p^n\vp(z) \vp(\om)\big\rangle \delta^{(2)}(z-\om) =(-1)^n \frac{2b^2}{n(n+1)} 
 \xi^{(n)}(\om),
\label{A13}
\ee
reproducing \rf{29}.

It is also easy to understand that \eq{29} remains valid also for the four-derivative action
(as well as for yet higher derivative actions) if the exponential is defined as the normal product.
This is because $\vp(z)$ pairs with only one $\vp(\om)$ modulo the renormalization of $\alpha$.
Among three parameters $b$, $q$ and $\alpha$ only two are independent, so
 we can set $\alpha=1$  to be nonrenormalized. 
The coefficient in front of the first term on the right-hand side 
of \eq{29} represents the conformal weight
of $\e^{\alpha \vp}$ and we arrive at \eq{DDK1}.
We have thus argued it remains unchanged for the higher-derivative action.

\tit{Central charge}

The central charge $c^{(\vp)}$ of $\vp$ can be computed from the variation of $T_{zz}$ under the infinitesimal conformal transformation  generated by \rf{hatdel} as
\be
 \LA \hat \delta T_{zz} (\om)\RA = \frac{ c^{(\vp)}} {12}\xi'''(\om)
\ee 
with the normal ordering in $T_{zz}$.

\paragraph{{\bf Quadratic action:}}

The energy-momentum tensor for the quadratic action  given by \rf{qTzz} with $\eps=0$.
 transforms as
\bea
\LA \hat\delta_\xi  T_{zz}^{(2)} (\om) \RA&=&\frac1{2b^2}  \int \d^2 z  
\big\langle q^2 \xi'''(z)+
 \xi'(z) \p^2 \vp(z)\vp(\om) +\xi(z) \p^3 \vp(z)\vp(\om) \big\rangle 
 \non && \hspace*{-1cm}\times  \delta^{(2)}(z-\om)
= 
 \frac {\xi'''(\om)}{2} \left(\frac{q^2}{b ^2} +\frac13-\frac 16 \right)= 
{\xi'''(\om)}\left( \frac{q^2}{2b ^2}+\frac 1{12}\right) , 
 \label{vp2}
\eea
where we have used \eq{A13}. The propagator here is {\it exact}\/ as before in \eq{32}:
this is why  $b^2$ cancels.
 In~\eq{vp2} $1/12$ corresponds to the usual quantum addition $1$ to the central charge.
The right-hand side of \eq{vp2} reproduces the DDK formula for the central charge.

\paragraph{{\bf Four-derivative action:}}

For the four-derivative action \eq{vp2} remains unchanged 
for the quadratic part of $T_{zz}$. The variation of the part
which is $\propto\!\eps$ reads for $G=0$~\cite{Mak23c}
\bea
\LA \hat\delta T_{zz}^{(4)} (\om)\RA_{G=0}
&=&\frac1{b^2} \int \d^2 z
 \LA 
\big[2 q \eps \xi'''(z) \p\bp\vp(z)
+(4q- 2)\eps \xi''(z) \p^2\bp \vp(z)  \right. \non &&
 \left. -6\eps \xi'(z) \p^3\bp \vp(z)
 -4\eps \xi(z) \p^4\bp \vp(z)\big]\vp(\om) \RA
\delta^{(2)}(z-\om)  \non  && 
=\frac {\xi'''(\om) }4 \left( -2 \cdot2 q+(4q- 2)\cdot 1+6 \frac 23 -4 \frac 12\right) =0,
\label{vp4}
\eea
where for the singular product  we have used~\cite{Mak22c} 
\be 
\int_{D_1}\!\d^2 z \, \xi(z) 
\big\langle  [-4\eps\p^{n+1}\bp\vp(z) \vp(\om)]\big\rangle \delta^{(2)}(z-\om) 
=(-1)^n \frac{2b^2}{(n+1)}  \xi^{(n)}(\om).
\label{A14}
\ee

The derivation of \eq{vp4} is similar to  \eq{32}.  
Only the average of the quadratic in $\vp(\om)$ part of $\delta T_{zz}$ is nonvanishing 
for the normal ordered $T_{zz}$: one  $\vp(\om)$ is annihilated by the variational derivative $\delta/\delta \vp(z)$, 
one more  $\vp(\om)$  pairs with $\p\vp(z)$. The renormalization of the
quadratic in $\vp$ part of $T_{zz}$  is taken into account  by the change $b_0^2\to b^2$ in \eq{qTzz}.

The consequence of \eq{vp4} is that
the central charge of $\vp$ coincides at $G=0$ (i.e. for the usual Liouville action plus $\eps R^2$)
with the one for the quadratic action. Moreover, this universality holds for the terms $\eps^2$ 
as well~\cite{Mak23c} and
most likely for higher terms in the DeWitt-Seeley expansion of the heat kernel and thus for the
Polyakov string.

We can repeat the computation for the part of $T_{zz}$ in \eq{qTzz} which involves $G$.
Again only the average of the quadratic in $\vp(\om)$ part of $\delta T_{zz}$ is nonvanishing 
for the normal ordered $T_{zz}$.
This makes the computations pretty much similar to those at one loop.
However, 
the contribution of higher loops is effectively  taken into account by the deviations of
 $b^2$, $q$  from there bare values. This is why I call it an ``intelligent'' one loop.

The contribution of the local part of $T_{zz}$ to $c^{(\vp)}$ vanishes like \rf{vp4}:
\be
\frac{1}{b^2}Gq\eps \int \d^2 z
\LA \big[-\xi'''(z)\p\bp\vp(z)
-2 \xi''(z) \p^2\bp \vp(z) \big] \vp(\om)
\RA \delta_\eps^{(2)}(z-\om) 
 = 0,
\label{tttp}
\ee
while for the nonlocal term we obtain
\bea
\lefteqn{\LA \hat \delta_\xi \left(-\frac{1}{b^2}Gq\eps 
\frac 1\bp \p^2 (\bp\vp \e^{-\vp} \p\bp \vp)\right)\RA } \non 
&&~= 
-\frac{1}{b_0^2}Gq\eps \int \d^2 z\, \xi'''(z)
 \LA\p\bp\vp(z)\vp(\om) \RA 
\delta^{(2)}(z-\om) = \frac 12 Gq \xi'''(\om) .~
\label{tttnlp}
\eea
In both cased we used \eq{A14}.
In contrast to the  classical analysis, now a non\-vanishing finite contribution arises
from the singular product.
Its occurrence is like in the CFT technique where it has arisen~\cite{Mak22} 
at one loop from the nonlocal (last) term in \rf{qTzz}.

Summing \rf{vp2} and \rf{tttnlp}, we obtain the central charge $c^{(\vp)}$ of $\vp$
shown in \eq{DDK2}. We have thus proved that it is exact~\cite{Mak23e}.
The vanishing of the total central charge is represented 
($d$ comes from $X^\mu$ and $-26$ comes from the ghosts) by the equation
\be
d-26+c^{(\vp)}=0.
\label{DDK2t}
\ee

\section{Relation to the minimal models}

It is easy to solve Eqs.~\rf{DDK1}, \rf{DDK2}, \rf{DDK2t} which modify KPZ-DDK for the four-derivative action \rf{inva}, adding a parameter $G$. The result reads
\bea
b^{-2} &=& 
\frac {13-d-6 G+\sqrt{(d_+-d)(d_--d)}}{12}, 
\label{tb}\\
d_\pm&=&13-6 G\pm 12 \sqrt{1+G} 
\label{dmp}
\eea
and $\gamma_{\rm str}$ is as in \eq{mygstr} with 
the KPZ barriers thus shifted to $d_\pm$ given by \rf{dmp}.

Their values depend on $G$ which has to lie in the interval $[-1,0]$ for the  
 the action~\rf{inva} to be stable as it follows from the identity (modulo boundary terms)
\be
\int \e^{-\vp} \left[(\p\bp \vp)^2-G \p\vp\bp\vp\p\bp\vp\right]=
\int\e^{-\vp} \left[(1+G)(\p\bp \vp)^2-G \nabla\p\vp\bar\nabla\bp\vp\right].
\ee
Then $b^{-2}$ is real for $d<d_-$ which increases from 1 at $G=0$ to 19 at $G=-1$.

As in the case of the usual Liouville theory the operators 
\be
V_\alpha=\e^{\alpha \varphi}, \qquad \alpha=\frac{1-n}2+\frac{1-m}{2b^2}
\ee
are the BPZ null-vectors for integer $n$ and $m$.
Their conformal weights 
$
\Delta_\alpha =\alpha+(\alpha-\alpha^2) b^2
$
coincide with Kac's spectrum of CFT with the central charge
$ 
c=1+6(b+b^{-1})^2
$
of the Virasoro algebra (not to be confused with the central 
charge $c^{(\vp)}$ of $\vp$) where $b$ is displayed in \eq{tb}.

To describe minimal models, we choose like in the usual Liouville theory
\be
c=25+6 \frac {(p-q)^2}{pq}, \quad G=\frac {(26-d-c)q}{6(q+p)},\quad
1-6\frac{(p-q)^2}{pq}\leq d\leq 19-6\frac pq
 \label{c25}
\ee
with coprime $q>p$. In that case this implied $d=26-c=1-6 \frac {(p-q)^2}{pq}$ for
the central charge of matter but now $d$ is 
a free parameter. 
From~\eq{tb} we find  
\be
b^{-2}=\left\{ 
\begin{array}{ll}
\displaystyle{\frac qp}&\hbox{perturbative branch}\nonumber\\[4mm]
\displaystyle{-1+\frac{(25-d)p}{6(q+p)} }\quad &\hbox{the other branch}\end{array} 
~~{\rm for}  
~d>25 -6\displaystyle{\frac {(p+q)^2}{p^2}}
\right.\!.
\ee
The perturbative branch is as in the usual 
Liouville theory, but the second branch is no longer 
$p \leftrightarrow q$ with it. There are no obstacles against $d=4$ for $q=p+1$!

\section{Conclusion}

The four-derivative term in the emergent action~\rf{inva} for $\vp$ revives,
telling the Nambu-Goto and Polyakov strings apart.
For the Polyakov string the higher-derivative terms do not change
the KPZ-DDK results (the universality holds).
The four-derivative Liouville theory is conformal invariant in spite of the presence of the
dimensionful worldsheet cutoff $\eps$ and
is exactly solved by the method of singular products.
The results are like KPZ-DDK except for
the barriers are shifted for the four-derivative action which may be also the case
for the Nambu-Goto string.
 All that is specific to the theory with diffeomorphism invariance.

I am indebted to the Organizing Committee for the opportunity to participate 
in the Conference and to meet 
old friends. My warmest Congratulations to Gordon on his Jubileum!



\begin{thebibliography}{10}

\bibitem{KPZ88} 
  V.G.~Knizhnik, A.M.~Polyakov and A.B.~Zamolodchikov,
  {\it Fractal structure of 2D quantum gravity,}
  Mod.\ Phys.\ Lett.\ A {\bf 3} (1988)  819.
\bibitem{Dav88} 
F.~David,
{\it Conformal field theories coupled to 2D gravity in the conformal gauge,}
  Mod.\ Phys.\ Lett.\ A {\bf 3} (1988)  1651.
\bibitem{DK89} 
  J.~Distler and H.~Kawai,
  {\it Conformal field theory and 2D quantum gravity,} 
  Nucl.\ Phys.\ B {\bf 321} (1989) 509.
\bibitem{Pol87}
A.M.~Polyakov, {\it Gauge fields and strings},  
Harwood Acad.\ Pub. (1987).

\bibitem{Mak23a}
Y.~Makeenko,
{\it Pauli-Villars' regularization of ghosts in path-integral string formulation}, 
JHEP {\bf 05} (2023)  085
[arXiv:2302.01954 [hep-th]]. 

\bibitem{Mak23c} 
Y.~Makeenko,
{\it Singular products and universality in higher-derivative conformal theory},
JHEP {\bf 09} (2023)  086 [arXiv:2307.06295 [hep-th]].

\bibitem{Mak21} 
  Y.~Makeenko,
  {\it Private life of the Liouville field that causes new anomalies in the Nambu-Goto string,}
  Nucl.\ Phys.\ B {\bf 967} (2021) 115398
  [arXiv:2102.04753 [hep-th]].
  
   \bibitem{DJ95}
 S. Deser and  R. Jackiw,
{\it Energy-momentum tensor improvements in two dimensions},
Int.\ J.\ Mod.\ Phys.\ B {\bf 10} (1996) 1499 
[arXiv:9510145 [hep-th]].

\bibitem{Mak22}
Y.~Makeenko,
{\it Opus on conformal symmetry of the Nambu-Goto versus Polyakov strings},  
Int. J. Mod.\ Phys.\ A {\bf 38} (2023) 2350010
[arXiv:2204.10205   [hep-th]].

\bibitem{Mak22c}
Y. Makeenko,
{\it Notes on higher-derivative conformal theory with nonprimary energy-momentum tensor
 that applies to the Nambu-Goto  string}, JHEP {\bf 01} (2023) 110 [arXiv:2212.02241 [hep-th]].

\bibitem{Mak23e}
Y.~Makeenko,
{\it Exact solution of higher-derivative conformal theory and minimal models},
Phys.\ Lett.\  {\bf 845} (2023) 138170  [arXiv:2308.05030 [hep-th]].

\end{thebibliography}
\end{document}